\documentstyle[aas2pp4]{article}

\let\oldfootsep=\footnotesep
\def\etal{et~al.}

\def\spose#1{\hbox to 0pt{#1\hss}}
\def\simlt{\mathrel{\spose{\lower 3pt\hbox{$\mathchar"218$}}
     \raise 2.0pt\hbox{$\mathchar"13C$}}}
     \def\simgt{\mathrel{\spose{\lower 3pt\hbox{$\mathchar"218$}}
  \raise 2.0pt\hbox{$\mathchar"13E$}}}

\begin{document}

\title
{The magnification invariant of simple galaxy lens models}
\author{N.~Dalal\\  
Physics Dept., University of California, San Diego, 
La Jolla, CA 92093 USA}
\begin{abstract}
We demonstrate that for several of the gravitational lens models used to 
describe galaxies, there exists a quantity we dub the magnification invariant,
equaling the sum of the signed magnifications of the images, that is
a constant when the image multiplicity is a maximum.  This invariant is 
independent of most of the model parameters, and is independent of the source 
position as long as the source lies inside the caustic.  It is not necessary 
to solve the lens equations to compute this invariant.  For quad lenses, it is 
usually easy to assign image parities and thus one can compute the sum of the 
signed fluxes, compare with the magnification invariant of different models, 
and infer the model magnification factors of the images without fitting.  
We evaluate the invariant for simple galaxy models, apply it
to known cases of quadruple lenses, and discuss implications for the 
ability of these lens models to reproduce observed image brightnesses.
\end{abstract}
\vspace{-5mm}
\section{INTRODUCTION}
Gravitational lensing is a well-known phenomenon that has
emerged as a useful tool in modern astrophysics.  Two important classes
of gravitational lenses are galaxy lenses, which can produce multiple
images of the background source with typical image separations of arcseconds 
(Keeton and Kochanek, 1996), and lensing by compact objects in our own
and nearby galaxies (eg Alcock \etal, 1997 and Alcock \etal, 1998).  The
latter class, termed ``microlenses'' since typical image separations are
too small to permit individual image resolution, are modeled as point masses.  
Models of galaxy lenses can be much more complicated, since the mass 
distributions of galaxies, although not well understood, are certainly far 
more complex than those of the starlike microlenses.  Since galaxies are 
thought to be dominated by dark halos in roughly
spherical, isothermal distributions, the first and most important part of 
a galaxy lens model is an isothermal term.  Variations on this can 
add ellipticity and/or shear into the mass distribution, or into the lens
potential itself (Kochanek, 1991; Kassiola \& Kovner, 1993; Kormann \etal, 
1994; Keeton \etal, 1997; 
Witt and Mao, 1997; Keeton and Kochanek, 1998).  The latter set of models
are generally much easier to deal with analytically, so we will concentrate
on those.

\cite{wm1} showed that for a binary microlensing system,
when the image multiplicity is a maximum (five in this case), then the
sum of the signed magnifications of all 5 images is always unity.  
The sign of the magnification of an image is merely the parity of the image, 
+1 for ordinary images and -1 for inverted images.  (see 
Schneider, Ehlers \& Falco (SEF), 1992 for a thorough review of lens theory 
and terminology.)  Writing $\mu_i$ as the
signed magnification of the $i^{th}$ image, let us define $I\equiv\sum_i\mu_i$.
This is a different quantity from the total magnification, which is given by
$\sum_i|\mu_i|$.  Witt \& Mao showed that $I=1$, independent of
quantities such as the Einstein radii of the lenses or the position of the
source, as long as the source was inside any of the caustics.  
Witt \& Mao ingeniously applied resultant theory to the
lens equations to obtain this result.  More recently, \cite{rhie} used 
a coordinate transformation to derive the same result, $I=1$, much more 
easily, and to show that $I=1$ for N point masses in a plane acting as a 
gravitational lens, for all nonnegative N.  Thus, for microlensing, the 
quantity $I$ is an invariant, independent of lens model, so we will refer to 
it as the ``magnification invariant''.  The invariance of the sum of the 
signed magnifications of the maximum number of images is useful for 
microlensing because, even though individual images cannot be resolved, the
total magnification is measurable, and conditions upon the sum of the
signed magnification imply conditions upon the sum of the unsigned 
magnifications, as Witt \& Mao describe.  Galaxy lensing presents us with the
reverse situation -- although the total magnification cannot be determined
observationally, individual images can be resolved.  Since for many lenses,
it is easy to assign image parities, one can compute the sum of the signed 
fluxes.  Therefore, if for galaxy lensing there are constraints upon the sum
of the signed fluxes, as there are for microlensing, these constraints 
should be observationally verifiable.  It is therefore interesting to ask 
whether such constraints exist
for models of galaxy lenses.  In this letter, we consider several of the 
commonly used models for galaxy lenses, and show that for many of them, 
although not all, such a property is true.  We then apply this knowledge
to several observed lens systems, and show how the magnification invariant
can be used to determine properties of the lens, without fitting.

\section{METHOD}
The method we employ is straightforward, and is based upon that used by
\cite{wm1} for the binary point lens.  For the simple lens potentials
we consider, we write the lens equations as polynomials in the variables 
${\vec x}$ and $\mu$, where ${\vec x}$ is the angular position and $\mu$ is 
the signed magnification.  We then apply elimination theory to eliminate the 
image positions and obtain a single equation satisfied by all of the 
magnifications.  Elimination theory is a well developed field within algebraic 
geometry, and there are multiple techniques and algorithms known to eliminate 
variables from simultaneous polynomial equations.  The most popular method is 
to compute the Gr\"obner basis of the original set of polynomials.  When 
lexicographic ordering of the variables is chosen, successive basis 
elements will have variables eliminated (Cox \etal, 1997).  A more classical 
technique is to compute the generalized resultant.  Sylvester, in the 19th
century, developed a formula to take the resultant of three polynomials in
three variables (Gelfand \etal, 1994), which is the category into which the 
lens equations fall.  Whatever technique is used, the result is a 
single polynomial equation in the magnification, whose roots are the 
signed magnifications of the images.  Thus, the degree of the final polynomial
equals the maximum number of images.  We should remark that it is not always
possible to perform such a procedure for all lens models, and it is important 
to verify that the degree of the final polynomial does equal the maximum
number of images.  Since the sum of the roots of a polynomial 
equals the ratio of the second coefficient to the leading coefficient (up to
a sign), to compute the sum of the magnifications, we need only 
compute this polynomial.  This ratio of coefficients then gives us the 
magnification invariant.  It is not even necessary to solve the lens equations!
This technique was first set out by Witt and Mao in their 1995 paper on binary
point lenses.

\section{LENS MODELS}
We now apply this technique to several popular galaxy lens models.  Galaxy
lens models, as stated earlier, generally consist of variations on the 
singular isothermal sphere (SIS).  The lens potential for an ordinary SIS
is $\psi=br$, where $b$ is the Einstein radius (which sets the physical scale
of the problem) and $r=\sqrt{x^2+y^2}$ is the distance to the center of the 
lens (SEF, 1992).  Galaxies are not perfectly spherical, however, and thus 
ellipticity and/or shear terms are added to the lens potential to account for 
this (see Blandford \& Kochanek, 1987; Kochanek, 1991; Kassiola \& Kovner, 
1993; Witt \& Mao, 1997).  For the models we consider, we let $\gamma$ 
measure the strength of the shear, and $\epsilon$ be the ellipticity.  
We choose without
loss of generality coordinates centered on the lens, and oriented along the 
ellipticity axis or direction of shear.  The models are :
\medskip

\noindent 1. SIS + elliptical : $\psi=br+\gamma br\cos(2 \theta)$ 

\noindent 2. SIE : $\psi=bR=b\sqrt{(1-\epsilon)x^2+(1+\epsilon)y^2}$

\noindent 3. SIS + external shear: $\psi=br+{\gamma\over 2}r^2\cos(2\theta)$

\noindent 4. SIE + external shear:
$\psi=bR+{\gamma\over 2}r^2\cos 2(\theta-\theta_\gamma)$

\medskip
\noindent Here, SIE stands for ``singular isothermal ellipse''.  This is
not the isothermal elliptical mass distribution described by \cite{kk93} and 
\cite{ksb}, but is instead an elliptical potential, corresponding to the 
singular isothermal elliptical potential (SIEP) of \cite{kk93}.
In model 4, the parameter $\theta_\gamma$ describes the orientation of the 
external shear relative to the ellipticity axes.  We consider here only 
singular models since lensed systems generally consist of even numbers of 
images (Kochanek, 1991), but later we will discuss one nonsingular model.

As an example of the procedure, we calculate $I$ for model~1, SIS+elliptical.  
For convenience, we use an orthonormal basis 
$\{e_{\hat r},e_{\hat\theta}\}=\{{\partial\over{\partial r}}, 
{1\over r}{\partial\over{\partial\theta}}\}$.
In polar coordinates, with the source at $(s,\theta_s)$ and image at 
$(r,\theta)$, the time delay becomes 
$$\tau={1\over 2}(r^2+s^2-2rs\cos(\theta-\theta_s))-br-\gamma br\cos(2\theta)$$

\noindent Stationarity of the time delay gives us the lens equations, the
solutions of which are the image positions:
$$\tau_{,\hat r}={{\partial\tau}\over{\partial r}}=r-s\cos(\theta-\theta_s) 
-b(1+\gamma\cos(2\theta))=0$$
$$\tau_{,\hat\theta}={1\over r}{{\partial\tau}\over{\partial\theta}}
=s \sin(\theta-\theta_s)+2\gamma b\sin(2\theta)=0$$

\noindent Note that the ${\hat\theta}$ equation is independent of $r$ 
(Kassiola and Kovner, 1995).  The components of the Hessian are
$$\tau_{;{\hat i}{\hat j}}=\left(\begin{array}{cc}
1&0\\ 0&1-{b\over r}+3\gamma {b\over r}\cos(2\theta)
\end{array} \right).$$

\noindent where we use semicolons to denote covariant derivatives.  Thus, 
the Jacobian determinant for this lens (see appendix) is 
$$\|J\|=1-{b\over r}+3\gamma {b\over r}\cos(2\theta),$$ and of course the
signed magnification is $\mu=1/\|J\|$, which is to be evaluated at each image 
position (i.e. at all of the solutions of the lens equations).

We now write this equation, and the two lens equations, as polynomial
equations.  Let $u=e^{i\theta}$ and $z=e^{i\theta_s}$.  Then the lens 
equations become
\begin{eqnarray}
&\gamma bu^4+{s\over z}u^3+2(b-r)u^2+szu+\gamma b=0,&\nonumber\\
&\gamma bu^4+{s\over{2z}}u^3-{{sz}\over 2}u-\gamma b=0,&\nonumber\\
&3\gamma b\mu u^4+2(\mu r-r-b\mu)u^2+3\gamma b\mu=0.\nonumber&
\end{eqnarray}

\noindent Eliminating $r$ and $u$, and assuming $\gamma\neq 0$, we obtain 
the desired fourth degree polynomial in $\mu$, which is too long to print 
here.  Again, this polynomial must be
fourth degree since there are at most four images.  We note that the 
leading coefficient, $a_4$, and the cubic coefficient, $a_3$, are related by

\begin{eqnarray}
a_4=-a_3&=&
{\frac{16{b^4}{{\gamma }^4}}{{z^4}}}
     ( 3{b^2}{s^4}(9-2{z^4}+9{z^8})
          {{\gamma }^2} \nonumber \\&&- {s^6}{z^4} 
- 768{b^4}{s^2}{z^4}{{\gamma }^4} + 
       4096{b^6}{z^4}{{\gamma }^6}) ,\nonumber
\end{eqnarray}

\noindent and so the sum of the roots of this polynomial equals 
$-a_3/a_4=1$.  This fourth 
degree polynomial has either 2 or 4 real roots.  When the source is inside the 
caustic, all roots are real and thus $I=\sum_i \mu_i=1$.  When the source is 
outside the caustic, 2 roots merge and become a complex conjugate pair.  The 
real roots correspond to visible images, while the complex roots correspond to 
spurious solutions to the lens equations, which are not visible images 
(Petters, 1993).  Therefore, when the number of real images is the maximum
possible (4 in this case), $I=1$.

Applying this procedure to all of the galaxy lens models, we obtain Table 1.
We should note that model 4 has the surprising property that when the shear is
oriented along the ellipticity axis (i.e. $\theta_\gamma=0$) and 
$\gamma=\epsilon$, then the caustic shrinks to a single point, located behind
the lens.  Although there are only two images for this special case, the two
magnifications sum to give the quantity listed in Table 1.

\begin{table}
\caption{Magnification invariants for the models shown.  Here, 
$r=\sqrt{x^2+y^2}$, $\theta=\tan^{-1}({y/x})$, 
$R=\sqrt{(1-\epsilon)x^2+(1+\epsilon)y^2}$.}
\begin{tabular}{lcc}
\hline
Model & $\psi$ & $\sum_{i}\mu_i$\\
\hline
0 SIS & $br$ & 2\\
1 SIS + elliptical & $br+\gamma br\cos(2 \theta)$ & 1\\
2 SIE & $bR$ & 2\\
3 SIS + external & $br+{\gamma\over 2}r^2\cos(2\theta)$ & 
${2\over{1-\gamma^2}}$\\
4 SIE + external & $bR+{\gamma\over 2}r^2\cos 2(\theta-\theta_\gamma)$ & 
${2\over{1-\gamma^2}}$\\
\hline
\end{tabular}
\end{table}

\section{DISCUSSION}
We have shown that for several models commonly used to describe galaxy lenses,
the predicted image magnifications must obey the condition that their sum
equal a simple constant, independent of source position (while the source lies
within the caustic) and often independent of model parameters.  This has some
interesting consequences.  To begin, we note that it is easy to compute the 
value of this
magnification invariant for a quad lens system, since the parities of the
images alternate as one goes around the ring of images.  Now consider, for
example, the object CLASS~1608+656 (Keeton \& Kochanek, 1996; Myers \etal, 
1995), which is a promising candidate for measurement of the Hubble constant
$H_0$.  By inspection, we
see that images A and B should have positive parity and that C and D should 
have negative parity (using the naming scheme used by Keeton \& Kochanek). 
Relative to B, A has brightness 2.06$\pm$0.06, C has 0.85$\pm$0.03, and
D has 0.26$\pm$0.03.  Thus, the magnification invariant for this system
is 1.95$\pm$0.07, multiplied by image B's unknown magnification factor.  Now,
Seitz \& Schneider (1994), applying the Raychaudhuri equation (see MTW, 1973 
or Wald, 1984), showed that minima of the time delay are never demagnified,
that is $\mu\ge 1$.  Since B is a minimum, its magnification factor must be 
$\ge 1$.  Therefore, we immediately see that model 1 cannot possibly reproduce 
CLASS~1608+656, since to do so would require that image B have magnification 
0.51$\pm$0.02.  The other models considered here are not expressly forbidden, 
but are quite unlikely to succeed, since they would require image B to have a 
magnification factor barely above the minimum possible value.  Thus we see, 
without the effort of modeling, that none of the simple models considered here 
have hope of recovering the positions and brightnesses of object 
CLASS~1608+656.  One can either ignore the fluxes when fitting, if their 
accuracy is suspect (which should not be the case for CLASS~1608+656, as 
fluxes were measured in the radio), or one must resort to more complicated 
models.  The problem, as Kochanek (1991) has emphasized, is that the 
complexity of model is limited by the small number of observables, which is
only 9 for a quad lens.  Note that the simple models here already have 5 or 7
parameters, so there are very few parameters that can be added before the
problem becomes underconstrained.  Of course, these remarks apply only for 
fitting on the image plane.  When fitting on the source plane, the images are 
solutions to different lens equations, and thus the sum of their 
magnifications need not be anything in particular.  Indeed, when images are 
near the critical curve, the sum of the magnifications can take on arbitrary 
values if the images do not have the same source, and so the magnification 
invariant would have little meaning.  However, for purposes other than 
verifying that an object is in fact gravitationally lensed, one ususally 
fits in the image plane.

Now, CLASS~1608+656 is a particularly difficult object to model, but what
about other lensed systems?  Consider the ``Cloverleaf'', H1413+117
(Keeton \& Kochanek, 1996; Yun \etal, 1997).  Using the $I$ band fluxes
listed by Keeton \& Kochanek, which should be the most trustworthy, we find
that the magnification invariant equals 0.07 times A's magnification factor.  
We shouldn't trust this very far however, since the invariant computed from
the $R$ band fluxes would be negative!  In any case, fitting any of the simple
models to H1413+117 would require very large image magnification factors
for a good fit, about 15 for model 1 or about 30 or higher for the others.  
This is using the $I$ band data; using the $R$ or $B$ data would give an even
smaller invariant, and therefore require even greater image magnifications.  
Since the fluxes of the four images are all comparable,
this means that {\it all} of the images would be required to lie very close to 
the critical curve, implying a tiny caustic with the source inside.  
Since the size of the caustic is determined by the magnitude of the 
shear/ellipticity, constraining the caustic to be small limits the range of
these parameters significantly.  Indeed, even though H1413+117 would seem
to be an ideal object for fitting -- one might go so far as to call it a 
``model'' lens system -- we were unable to fit any of the listed models
satisfactorily to it, using the $\chi_{\rm S}^2$ error function described in
Kochanek (1991), except for model 4.  By satisfactory fit, we mean 
$\chi^2/N_{dof}\simeq 1$.  Model 4 was able to give a good
fit due to the feature mentioned above, that when the external shear and 
ellipticity are related in a specific way, the caustic shrinks to a point,
even though the ellipticity may be significant.  This feature enables model 4 
to attain the large magnifications necessary, while possessing the ellipticity
required to fit the image geometry.  The other models, which cannot be
elliptical without having a large caustic, are thus doomed to failure.  Our 
difficulty in fitting this 
object is not much of a revelation; previous efforts at fitting H1413+117 
using simple models such as the ones discussed here are unable to recover the 
image brightnesses satisfactorily.  For example, neither \cite{kayser} 
nor \cite{koch1} attempted to fit the brightnesses.  \cite{jj} do
fit the flux ratios, using a variant of model 4 with extra parameters.  Their
model is similar enough to model 4 to possess the same ``shrinking caustic''
feature, and not surprisingly their best fit has a tiny (tangential) caustic
barely enclosing the source QSO, with the associated large image 
magnifications of order 50.  The prerequisite that the caustic be tiny 
for a good fit makes sense in terms of the magnification invariant.

H1413+117 also illustrates another use of the magnification invariant.  By 
merely adding up four numbers, we immediately learn that the image 
magnifications must be large.  Yun \etal\ point out that this means that the 
curvature of the time delay surface at the ring of images is small, meaning 
that the relative time delays between images must be small.  We have seen, 
however, that the magnifications must be large for all models, and therefore
the predicted relative time delays between images will be small, independent 
of model.  This is useful knowledge for $H_0$ determination.

We have so far discussed only singular models.  Lens potentials can be 
rendered nonsingular by the introduction of a core radius.  Altering the
models discussed here to make them nonsingular destroys the invariance 
of the sum of the magnifications.  However, the size of the core radius term 
is limited by the absence of a central image from observed lens systems.  For 
small enough core radius the sum of the magnifications of the remaining four 
images will be close to the value for the singular model.  Thus, our results 
should still apply for the cases of interest, where no central image is 
observed.  For models based upon elliptical mass distributions, one can see 
from numerical calculations that the sum of the  magnifications of the maximum 
number of images is not a constant.  However, for typical parameter values, 
such as in \cite{b0712}, \cite{m1608} or \cite{nair}, we find
that the sum of the magnifications of four images is between 2 and 3, so 
this model should be marginally better at fitting objects like CLASS~1608+656.
Models with multiple galaxies (e.g. Hogg \& Blandford, 1994)
can have a much higher magnification invariant, however they have a maximum 
number of images not equal to four. 

The magnification invariant, we have seen, is easy to calculate for observed
quad lenses, since one merely adds up four numbers.  Doing so gives us 
useful information on the ability of lens models to fit the observations, 
even before fitting.  We have seen that for small invariants, such as 
H1413+117, large magnifications are necessary.  On the other hand, invariants
that are large, such as for CLASS~1608+656, can help rule out models {\it a 
priori}.  It is interesting that two lenses recently discovered, B1933+503 
(Sykes \etal, 1997) and B0712+472 (Jackson \etal, 1998) both have large
invariants and should, like CLASS~1608+656, be hard to fit using the simple
models.  We hope that models based on realistic mass distributions in galaxies,
e.g. \cite{kkoch2}, willl fare better at describing these objects.

\section*{ACKNOWLEDGEMENTS}
The author would like to thank John Wavrik for discussions of elimination 
theory, Jeffrey Rabin and Roger Blandford for many helpful conversations, and 
Kim Griest for encouragement, discussions, and for reading the manuscript.  
This work was supported in part by the US DOE under grant DEFG0390ER 40546.

\appendix
\section*{APPENDIX}
In an arbitrary frame, it is not obvious how the Jacobian of the lens mapping
is related to the Hessian of the time delay.  We can easily relate these two 
by noting that in the usual Cartesian coordinates, 
$J^i_{\ j}=\tau^{;i}_{\ \ j}$, 
i.e. that the components of the Jacobian matrix equal the components of the 
Hessian.  Note, however, that the Hessian in this equation
has been contracted with the metric tensor, since the Hessian is a  
${0\choose 2}$-tensor while the Jacobian, a linear transformation taking 
1-forms to 1-forms, must be a ${1\choose 1}$-tensor.  Of course, in 
orthonormal frames such as the one used above, or Cartesian coordinates, the 
metric tensor is merely the identity matrix.  Knowing the relation between the 
Jacobian and Hessian in Cartesian coordinates tells us their relationship in 
arbitrary frames.  Let $dx^{\hat i}$ be the 1-form dual to the basis vector 
$e_{\hat i}$ in the image plane, and similarly for $ds^{\hat i}$ in the source 
plane.  Then
$$ds^{\hat i}={{\partial s^{\hat i}}\over{\partial s^j}}
{{\partial s^j}\over{\partial x^k}}{{\partial x^k}\over{\partial x^{\hat l}}}
dx^{\hat l}$$

\noindent where unhatted components are Cartesian, and hatted components are 
in arbitrary frames.  Substituting the Cartesian lensing Jacobian
$${{\partial s^j}\over{\partial x^k}}=\tau^{;j}_{\ \ k}=
{{\partial x^j}\over{\partial x^{\hat m}}}\tau^{;\hat m}_{\ \ \hat n}
{{\partial x^{\hat n}}\over{\partial x^k}}$$

\noindent into the previous equation gives us the lensing Jacobian expressed
in terms of the Hessian computed in an arbitrary frame.
$$J^{\hat i}_{\ \hat l}={{\partial s^{\hat i}}\over{\partial s^j}}
{{\partial x^j}\over{\partial x^{\hat k}}}\tau^{;\hat k}_{\ \ \hat l}.$$  

\noindent For orthonormal frames, the Jacobian matrices 
${{\partial s^{\hat i}}\over{\partial s^j}}$ and 
${{\partial x^j}\over{\partial x^{\hat k}}}$ are merely rotation matrices, with
unit determinant, so that the Jacobian determinant equals the determinant of
the Hessian matrix.

\eject
\end{document}